\documentclass[lettersize,journal]{IEEEtran}
\usepackage{amsmath,amsfonts}
\usepackage{algorithmic}
\usepackage{algorithm}
\usepackage{array}
\usepackage[caption=false,font=normalsize,labelfont=sf,textfont=sf]{subfig}
\usepackage{textcomp}
\usepackage{stfloats}
\usepackage{url}
\usepackage{amssymb}
\usepackage{verbatim}
\usepackage{multirow}
\usepackage{graphicx}
\usepackage{cite}
\usepackage[dvipsnames,usenames]{color}
\usepackage[usenames,dvipsnames]{color}
\newcommand{\algorithmcfname}{\textbf{Method 1 }} 

\begin{document}
	%
	\title{Rank Optimization for MIMO Channel with RIS: Simulation and Measurement}
	\vspace{-0.2cm}
	\author{
		Shengguo Meng, Wankai Tang, Weicong Chen, Jifeng Lan, 
		\\ Qun Yan Zhou, Yu Han, Xiao Li, and Shi Jin
		\vspace{-1.2cm}
		\thanks{
			\vspace{-0.6cm}
			\par This work was supported in part by the National Key Research and Development Program of China under Grant 2023YFB3811505, the National Natural Science Foundation of China under Grants 62261160576,  62231009, 61971126, 62201138, U22A2002, and 62301148, the Natural Science Foundation of Jiangsu Province under Grants BK20220809 and BK20230824, the China Postdoctoral Science Foundation under Grant BX20230065, and the Jiangsu Excellent Postdoctoral Program under Grant 2023ZB476. (\emph{Corresponding authors: Shi Jin and Wankai Tang.})
			\par Shengguo Meng, Wankai Tang, Weicong Chen, Jifeng Lan, Yu Han, Xiao Li, and Shi Jin are with the National Mobile Communications Research Laboratory, Southeast University, Nanjing 210096, China.  (e-mail: seumengsg@seu.edu.cn; tangwk@seu.edu.cn; cwc@seu.edu.cn; lanjifeng@seu.edu.cn; hanyu@seu.edu.cn; li\_xiao@seu.edu.cn; jinshi@seu.edu.cn).
			\par Qun Yan Zhou is with the State Key Laboratory of Millimeter Waves, Southeast University, Nanjing 210096, China. (e-mail: qyzhou@seu.edu.cn).
	}}
	\maketitle
	\vspace{-0.2cm}
	\vspace{-0.2cm}
	\begin{abstract}
		Reconfigurable intelligent surface (RIS) is a promising technology that can reshape the electromagnetic environment in wireless networks, offering various possibilities for enhancing wireless channels. Motivated by this, we investigate the channel optimization for multiple-input multiple-output (MIMO) systems assisted by RIS. In this paper, an efficient RIS optimization method is proposed to enhance the effective rank of the MIMO channel for achievable rate improvement. Numerical results are presented to verify the effectiveness of RIS in improving MIMO channels. Additionally, we construct a 2$\times$2 RIS-assisted MIMO prototype to perform experimental measurements and validate the performance of our proposed algorithm. The results reveal a significant increase in effective rank and achievable rate for the RIS-assisted MIMO channel compared to the MIMO channel without RIS.
	\end{abstract}  
	\vspace{-0.2cm}
	\begin{IEEEkeywords}
		Reconfigurable intelligent surface,  multiple-input multiple-output, measurement, 6G.
	\end{IEEEkeywords}
	%
	\IEEEpeerreviewmaketitle

	\vspace{-0.6cm}
	\section{Introduction}
	\par \IEEEPARstart{I}{n} recent years, the increasing presence of various mobile communication services such as video streaming and online gaming has sparked a significant surge in the need for higher data rates. To meet the growing demands, various technologies have been applied. Among them, multiple-input multiple-output (MIMO) technology stands out as a solution to increase system capacity without the need for additional frequency resources. However, as MIMO is widely deployed, new challenges have arisen. In propagation environments characterized by strong line-of-sight (LoS) path and limited scatterings, the correlation in MIMO channels increases. This phenomenon weakens the multiplexing of MIMO, resulting in a reduced capability to support multiple data streams and subsequently diminishing the performance of communication systems. Therefore, there is an urgent need for innovative solutions to optimize the effective rank of MIMO channels to address these challenges. The advent of reconfigurable intelligent surfaces (RIS), which can be integrated into the channel, offers new opportunities for addressing this problem.
	
	\par As one of the key potential technologies for the sixth-generation (6G) networks, RIS can dynamically manipulate electromagnetic waves \cite{Meta_Cui, RIS_M_Di, Han_2019, Sang_2022, Meng_2023}, thus holding the potential to customize wireless channels\cite{Chen_2022}. Some studies have been conducted on leveraging RIS as part of the MIMO channels to improve the achievable rate of the systems \cite{li_ergodic_2023, perovic_achievable_2021, chen_channel_2023}. These related works demonstrate the capability of RIS to improve achievable rates of MIMO systems in different scenarios. Furthermore, RIS can be employed to improve the system performance by directly enhancing the effective rank \cite{roy_effective_2007} or minimum singular value of the MIMO channel \cite{M.A._2021}. However, the optimization algorithms proposed in the aforementioned references are complex and may not be easily applicable to practical systems, rendering a lack of relevant experimental measurements.
	
	\par In this paper, we first develop an RIS-assisted MIMO channel model and then propose an efficient RIS optimization method called the maximum cross-swapping algorithm (MCA) to enhance the effective rank of the MIMO channel for achievable rate improvement. Numerical simulations are conducted to verify the effectiveness of RIS in improving the effective rank and the achievable rate. Moreover, we set up an RIS-assisted 2$\times$2 MIMO communication prototype to validate the performance of our proposed algorithm through experimental measurements. The results show that compared to the MIMO channel without RIS, the RIS-assisted MIMO channel, optimized using the MCA, achieves a notable increase of 30.1\% in the effective rank and 13.6\% improvement in the achievable rate.

	
	\begin{figure}[tbp]
		\centering
		\includegraphics[height=1.61in]{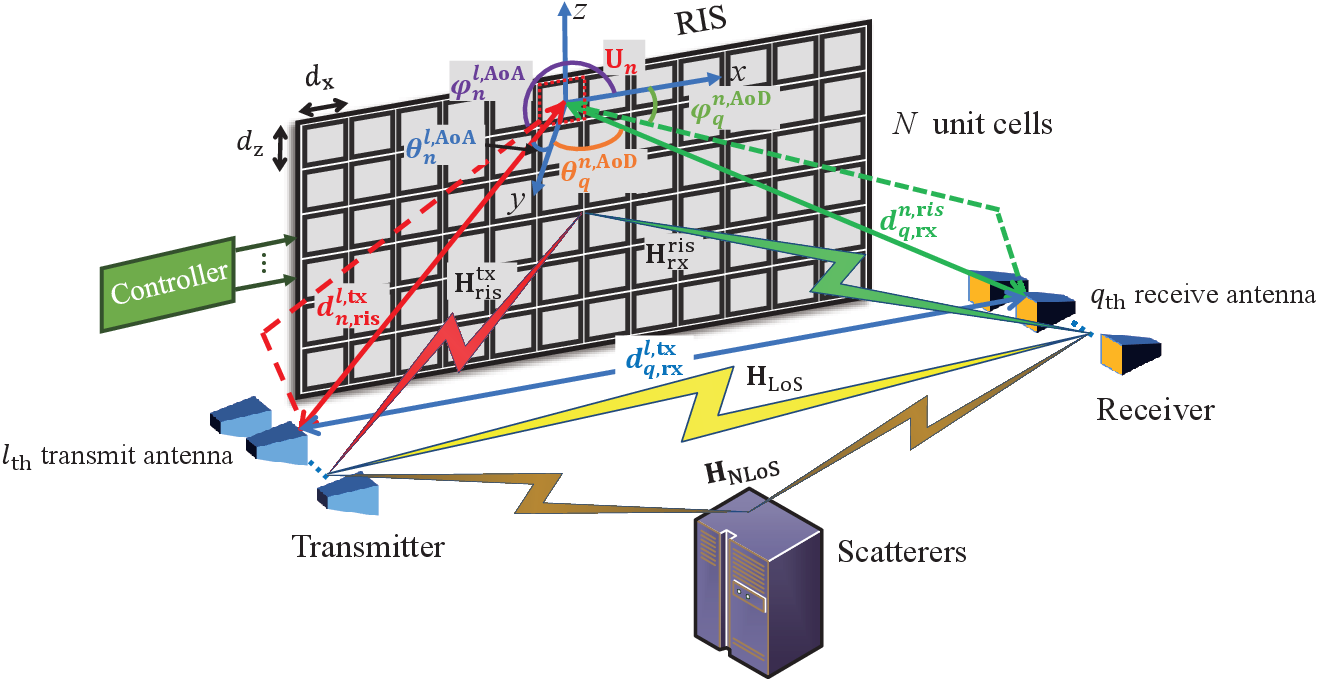}
		\vspace{-0.45cm}
		\caption{\centering The considered RIS-assisted MIMO communication system.}
		\label{Systemmodel}
		\vspace{-0.5cm}
	\end{figure}
	
    \vspace{-0.5cm}
	\section{System Model}
	We consider an RIS-assisted MIMO communication system as shown in Fig.\ref{Systemmodel}. The transmitter and the receiver have $L$ and $Q$ antennas, respectively. The RIS consists of $N$ unit cells. The entire MIMO channel between the transmitter and the receiver is comprised of the RIS-assisted channel and the non-RIS-assisted channel \cite{M.A._2021, Han_2019}, which can be modeled as 
	\vspace{-0.1cm}
	\begin{equation} \label{equ1}
		\begin{aligned}
			\mathbf{H}=&\sqrt{\alpha}\mathop {\underbrace{\mathbf{H}_{\mathrm{rx}}^{\mathrm{ris}}\mathbf{\Gamma H}_{\mathrm{ris}}^{\mathrm{tx}}}} \limits_{\mathrm{RIS-assisted\ channel}}\\
			&+\sqrt{1-\alpha}\left( \underset{\mathrm{non-RIS-assisted\ channel}}{\underbrace{\sqrt{\frac{K}{1+K}}\mathbf{H}_{\mathrm{LoS}}+\sqrt{\frac{1}{1+K}}\mathbf{H}_{\mathrm{NLoS}}}} \right) 
		\end{aligned},
		\vspace{-0.1cm}
	\end{equation}
	where $\mathbf{H}_{\mathrm{ris}}^{\mathrm{tx}}\in \mathbb{C} ^{N\times L}$ and $\mathbf{H}_{\mathrm{rx}}^{\mathrm{ris}}\in \mathbb{C} ^{Q\times N}$ represent the channel between the transmitter and the RIS, and the channel between the RIS and the receiver, respectively. $\mathbf{\Gamma} \in \mathbb{C}^{N\times N}$ is a diagonal matrix composed of the reflection coefficients of each unit cell. $\mathbf{H}_{\mathrm{LoS}}\in \mathbb{C} ^{Q\times L}$ and $\mathbf{H}_{\mathrm{NLoS}}\in \mathbb{C} ^{Q\times L}$ represent the non-RIS-assisted line of sight (LoS) and non-LoS (NLoS) paths between the transmitter and the receiver, respectively. $\alpha \in (0, 1)$ denotes the power ratio of the RIS-assisted channel, i.e., the power ratio of $\mathbf{H}_{\mathrm{rx}}^{\mathrm{ris}}\mathbf{\Gamma}\mathbf{H}_{\mathrm{ris}}^{\mathrm{tx}}$ in the overall channel $\mathbf{H}$. $K$ is the Rician factor of the non-RIS-assisted channel.
	
	\par Specifically, the channel between the transmitter and the RIS can be expressed as
	\vspace{-0.2cm}
	\begin{equation}\label{equ2}
		\mathbf{H}_{\mathrm{ris}}^{\mathrm{tx}}=\left[ \begin{matrix}
			h_{1,\mathrm{ris}}^{1,\mathrm{tx}}&		h_{1,\mathrm{ris}}^{2,\mathrm{tx}}&		\cdots&		h_{1,\mathrm{ris}}^{L,\mathrm{tx}}\\
			h_{2,\mathrm{ris}}^{1,\mathrm{tx}}&		h_{2,\mathrm{ris}}^{2,\mathrm{tx}}&		\cdots&		h_{2,\mathrm{ris}}^{L,\mathrm{tx}}\\
			\vdots&		\vdots&		\ddots&		\vdots\\
			h_{N,\mathrm{ris}}^{1,\mathrm{tx}}&		h_{N,\mathrm{ris}}^{2,\mathrm{tx}}&		\cdots&		h_{N,\mathrm{ris}}^{L,\mathrm{tx}}\\
		\end{matrix} \right],
		\vspace{-0.2cm}
	\end{equation}
	where $h_{n,\mathrm{ris}}^{l,\mathrm{tx}}$ is the channel coefficient between the $l_{\mathrm{th}} (l = 0, 1, \dots, L)$ transmit antenna and the $n_{\mathrm{th}} (n = 0, 1, \dots, N)$ unit cell $\mathrm{U}_{n}$. We assume that both the transmitter and receiver are in the near field of the RIS, but in the far field of each unit cell of the RIS. Then considering the wireless signal propagation loss between them, the channel coefficient can be modeled as \cite{tang_mimo_2020}
	\vspace{-0.2cm}
	\begin{equation}\label{equ3}
		h_{n,\mathrm{ris}}^{l,\mathrm{tx}}=\sqrt{\frac{G_\mathrm{tx} F\left( \theta _{n}^{l,\mathrm{AoA}},\varphi _{n}^{l,\mathrm{AoA}} \right) d_{\mathrm{x}}d_{\mathrm{z}}}{4\pi \left({d_{n,\mathrm{ris}}^{l,\mathrm{tx}}}\right) ^2}}\times e^{-j2\pi d_{n,\mathrm{ris}}^{l,\mathrm{tx}}/\lambda},
		\vspace{-0.2cm}
	\end{equation}
	where $G_{\mathrm{tx}}$ denotes the gain of each transmit antenna, $\theta_{n}^{l,\mathrm{AoA}}$ and $ \varphi_{n}^{l,\mathrm{AoA}}$ denote the angle of arrival (AoA) from the $l_{th}$ transmit antenna to the unit cell $\mathrm{U}_{n}$. $F(\theta,\varphi)$ is the normalized power radiation pattern of each unit cell\cite{RIS_path_1}. $d_{\mathrm{x}}$ and $d_{\mathrm{z}}$ denote the width and length of each unit cell, respectively. $d_{n,\mathrm{ris}}^{l,\mathrm{tx}}$ denotes the distance between the $l_{th}$ transmit antenna and the unit cell $\mathrm{U}_{n}$, and $\lambda$ represents the wavelength. In analogy with $\mathbf{H}_{\mathrm{ris}}^{\mathrm{tx}}$, $\mathbf{H}_{\mathrm{rx}}^{\mathrm{ris}}$ can be modeled.

	\par Furthermore, the reflection matrix of the RIS $\mathbf{\Gamma}$ can be represented as
	\vspace{-0.2cm}
	\begin{equation}\label{equ4}
		\mathbf{\Gamma }=\mathrm{diag}\left\{ e^{j\phi _1}, e^{j\phi _2}, \dots, e^{j\phi _N} \right\},
		\vspace{-0.2cm}
	\end{equation}
	where $e^{j\phi_{n}}$ denotes the reflection coefficient of unit cell $\mathrm{U}_{n}$. In practice, continuous control of the reflection phase for each unit cell incurs significant hardware implementation costs. Thus, we consider the discrete implementation, where the reflection phase is taken only from a finite number of discrete values. We assign $\pmb{\phi }=\left[ \phi _1, \phi _2, \dots, \phi _N \right] $ to be the phase configuration vector of the RIS, satisfying $	\forall \phi _n\in \Xi _b=\left\{ 0,\frac{2\pi}{2^b}, \dots, \frac{ 2\pi \left(2^b-1\right)} {2^b} \right\}$, where $b$ denotes quantization bit of reflection phase of each unit cell.
	
	\par As to the non-RIS-assisted channel, the LoS path between
	the transmitter and the receiver can be modeled as
	\vspace{-0.1cm}
	\begin{equation}\label{equ5}
		\mathbf{H}_{\mathrm{LoS}}=\left[ \begin{matrix}
			h_{1, \mathrm{rx}}^{1, \mathrm{tx}}&		h_{1,\mathrm{rx}}^{2, \mathrm{tx}}&		\cdots&		h_{1,\mathrm{rx}}^{L, \mathrm{tx}}\\
			h_{2,\mathrm{rx}}^{1, \mathrm{tx}}&		h_{2,\mathrm{rx}}^{2, \mathrm{tx}}&		\cdots&		h_{2,\mathrm{rx}}^{L, \mathrm{tx}}\\
			\vdots&		\vdots&		\ddots&		\vdots\\
			h_{Q,\mathrm{rx}}^{1, \mathrm{tx}}&		h_{Q,\mathrm{rx}}^{2, \mathrm{tx}}&	\cdots&		h_{Q,\mathrm{rx}}^{L, \mathrm{tx}}\\
		\end{matrix} \right],
	\end{equation}
	where $h_{q,\mathrm{rx}}^{l,\mathrm{tx}}=\sqrt{G_{\mathrm{tx}}G_{\mathrm{rx}}}\lambda /\left( 4\pi d_{q,\mathrm{rx}}^{l,\mathrm{tx}} \right) \times e^{-j2\pi d_{q,\mathrm{rx}}^{l,\mathrm{tx}}/\lambda}
	$ denotes the channel coefficient of the LoS path between the $l_{\mathrm{th}}$ transmit antenna and the $q_{\mathrm{th}} (q = 0, 1, \dots, Q)$ receive antenna, $d^{l,\mathrm{tx}}_{q,\mathrm{rx}}$ and  $G_{\mathrm{rx}}$ are the distance between them and the gain of each receive antenna, respectively. In addition, the NLoS path can be represented by 
	\vspace{-0.3cm}
	\begin{equation}\label{equ6}
		\mathbf{H}_{\mathrm{NLoS}}=\frac{\sqrt{G_{\mathrm{tx}}G_{\mathrm{rx}}}\lambda}{4\pi d_{\mathrm{rx}}^{\mathrm{tx}}}\tilde{\mathbf{H}}_{\mathrm{NLoS}},
		\vspace{-0.2cm}
	\end{equation}
	where $d_{\mathrm{rx}}^{\mathrm{tx}}$ is the distance between the center of the transmitter and the center of the receiver, and $\tilde{\mathbf{H}}_{\mathrm{NLoS}} \in \mathbb{C}^{L\times Q}$ is the NLoS component. Each element of $\tilde{\mathbf{H}}_{\mathrm{NLoS}}$ is an i.i.d. complex Gaussian random variable with zero mean and unit variance.
	
	\vspace{-0.3cm}
	\section{Optimization Method}
	In this section, we introduce an optimization method to improve the MIMO channel. The effective rank is used as a metric to evaluate the orthogonality among the MIMO subchannels\cite{M.A._2021}. According to the definition in \cite{roy_effective_2007}, the effective rank of the MIMO channel $\mathbf{H}$ between the transmitter and the receiver can be expressed as
	\vspace{-0.2cm}
	\begin{equation}		\label{equ7}		
		\mathrm{erank}(\mathbf{H})=\exp \left( \sum_{i=1}^{\mathrm{rank}(\mathbf{H})}{-\sigma _i^{\prime}\ln \sigma _i^{\prime}} \right),
		\vspace{-0.1cm}
	\end{equation}
	where $\mathrm{rank}(\mathbf{H})$ is the rank of the MIMO channel, and $\sigma _i^{\prime}={\sigma _i}/{\sum_i{\sigma _i}}
	$ is the normalized singular value.
	
	\par By incorporating RIS as a component of the MIMO channel, it is possible to optimize the configuration of the RIS $\pmb{\phi}$ to improve the effective rank of the MIMO channel $\mathbf{H}$, thus improving the orthogonality of the subchannels and overall system performance. Hence, the optimization goal can be written as
	\vspace{-0.3cm}
	\begin{equation}\label{equ8}
		\begin{aligned}
			&\underset{\pmb{\phi }}{\mathrm{maximize}}\,\,\mathrm{erank}\left( \mathbf{H} \right)\\
			&\,\,\mathrm{subject} \ \mathrm{to} \ \forall \phi _n\in \Xi _b,\ n=1,2,...,N.\\
		\end{aligned}
		\vspace{-0.1cm}
	\end{equation}
	
	\par We propose an efficient optimization method named MCA to optimize the phase configuration of the RIS. MCA is a codebook-based method, which selects superior configurations from the codebook and generates optimal configurations by cross-swapping parts of these superior configurations. The specific procedure of MCA is described as follows.
	
	\subsubsection{Iterating through random configurations} In the MCA, we first generate a set of $T$ random RIS phase configurations, i.e., $	\varPsi =\left\{ \pmb{\phi }_1, \dots,\pmb{\phi}_T \right\}$. Then, we configure the RIS according to these $T$ random configurations and obtain the corresponding effective rank of the MIMO channel $\mathbf{H}$. Furthermore, the $T$ random RIS configurations are sorted in ascending order based on their individual effective ranks, and the top two configurations are selected as the parent set $\varPsi_{\mathrm{parent}}$ to generate a new set of RIS phase configurations for the next round. The corresponding parent set is expressed as
	\vspace{-0.3cm}
	
	\begin{equation}\label{equ9}
		\varPsi _{\mathrm{parent}}=\left\{ \pmb{\phi }_{\mathrm{max}},\pmb{\phi }_{\mathrm{submax}} \right\}, 
		\vspace{-0.1cm}
	\end{equation}
	where $\pmb{\phi}_{\mathrm{max}}$ and $\pmb{\phi}_{\mathrm{submax}}$ represent the RIS configuration with the highest and second-highest effective rank of the corresponding MIMO channel $\mathbf{H}$ in the set $\varPsi$, respectively.
	
	\subsubsection{Cross-swapping the parent configurations} We generate offspring RIS phase configurations denoted as $\varPsi _{\mathrm{son}}=\left\{ \pmb{\phi }_{1}^{\mathrm{son}},\pmb{\phi }_{2}^{\mathrm{son}},\dots,\pmb{\phi}_{2^{N_{\mathrm{new}}}}^{\mathrm{son}} \right\}$ by sequentially cross-swapping the phase designs of the first  $N_{\mathrm{new}}$ unit cells from the parent configurations  $\left\{ \pmb{\phi }_{\mathrm{max}},\pmb{\phi }_{\mathrm{submax}} \right\} $. The detailed procedure is outlined in Method 1.
	\subsubsection{Obtaining the optimal offspring configuration} After applying each newly generated offspring configuration to the RIS, we can calculate the corresponding effective rank of the MIMO channel $\mathbf{H}$. The optimal offspring configuration, denoted as $\pmb{\phi}_{\mathrm{mca}}$, is determined as the one with the highest effective rank among the set $\varPsi_{\mathrm{son}}$. The whole process of the MCA is summarized in Algorithm \ref{ag2}. 
	\vspace{-0.2cm}
	\begin{algorithm}
		{\algorithmcfname Cross-swapping Parent Configurations to Generate Offspring Configurations.}
		\\{\noindent}	 \rule[8pt]{8.86cm}{0.05em}\\
		\label{ag1}
		\vspace{-23pt}
		\begin{algorithmic}[1]
			\STATE $\mathbf{input:}	\ N_{\mathrm{new}},\ \pmb{\phi }_{\mathrm{max}},\ \pmb{\phi }_{\mathrm{submax}}
			$
			\STATE $\mathbf{for} \  i= 1,2,\dots, 2^{N_{\mathrm{new}}} \ \mathbf{do}$
			\STATE \hspace{0.5cm} convert $i$ to a binary number $i_{\mathrm{b}}$ with $N_{\mathrm{new}}$ bits
			\STATE \hspace{0.5cm} $\mathbf{for} \ j = 1,2,\dots, N_{\mathrm{new}} \ \mathbf{do}$ 
			\STATE \hspace{1cm} 
			$\mathbf{if}$ the $j_{\mathrm{th}}$ bit of $i_{\mathrm{b}}$ equals to 0
			\STATE \hspace{1.5cm}	    
			assign the $j_{\mathrm{th}}$ element of $\pmb{\phi }_{\mathrm{max}}$ to the $j_{\mathrm{th}}$  \\ \hspace{1.5cm} element of $ \pmb{\phi }_{i}^{\mathrm{son}}$
			\STATE \hspace{1cm} 
			$\mathbf{else}$ $\mathbf{if}$ the $j_{\mathrm{th}}$ bit of $i_{\mathrm{b}}$ equals to 1
			\STATE \hspace{1.5cm}	    
			assign the $j_{\mathrm{th}}$ element of $\pmb{\phi }_{\mathrm{submax}}$ to the $j_{\mathrm{th}}$  \\ \hspace{1.5cm} element of $\pmb{\phi }_{i}^{\mathrm{son}}$
			\STATE \hspace{1cm}
			$\mathbf{end\ if}$
			\STATE \hspace{1cm} assign the remaining elements of $\pmb{\phi}_{\mathrm{max}}$ to the
			\\ \hspace{1cm} remaining elements of $\pmb{\phi }_{i}^{\mathrm{son}}$
			\STATE \hspace{0.5cm} $\mathbf{end}\ \mathbf{for}$
			\STATE $\mathbf{end\ for}$
			\STATE $\mathbf{output:}\ $offspring RIS configuration set  $\varPsi _{\mathrm{son}}=\left\{ \pmb{\phi }_{1}^{\mathrm{son}},\pmb{\phi }_{2}^{\mathrm{son}},\dots,\pmb{\phi}_{2^{N_{\mathrm{new}}}}^{\mathrm{son}} \right\}$ 
		\end{algorithmic}
	\end{algorithm}
	\vspace{-0.4cm}

	\begin{algorithm}
		\caption{Maximum Crosss-swapping Algorithm (MCA).}
		\label{ag2}
		\begin{algorithmic}[1]
			\STATE $\mathbf{input:} N, \ T, \ N_{\mathrm{new}}, \ b$
			\STATE generate a set $ \varPsi$ containing $T$ random RIS configurations $ \left\{ \pmb{\phi }_t \right\} _{t=1}^{T}$ based on $\Xi_{b}$
			\STATE calculate the effective rank of the MIMO channel $\mathbf{H}$ with the RIS configured by each configuration $\pmb{\phi}_{t}$
			
			\STATE find the RIS configuration $\pmb{\phi}_{\mathrm{max}}$ and $ \pmb{\phi}_{\mathrm{submax}}$ with the highest and second-highest effective rank of the corresponding MIMO channel $\mathbf{H}$ in the set $\varPsi$, respectively. 
			\STATE generate offspring RIS configuration set $\varPsi_{\mathrm{son}}$ according to \textbf{Method 1}
			\STATE 	$\mathbf{for} \  i= 1,2,\dots, 2^{N_{\mathrm{new}}} \ \mathbf{do}$
			\STATE \hspace{0.5cm} calculate the corresponding effective rank of the 
			\\ \hspace{0.5cm} MIMO channel $\mathbf{H}$ with the RIS configured by $\pmb{\phi}_{i}^{\mathrm{son}}$
			\STATE $\mathbf{end\ for}$
			\STATE $\mathbf{output:}$  $\pmb{\phi}_{\mathrm{mca}}=\underset{\pmb{\phi }\in \varPsi_{\mathrm{son}}}{\mathrm{arg} \max}\,\,\mathrm{erank}\left( \mathbf{H} \right)$
			\end{algorithmic}
	\end{algorithm}
	\vspace{-0.3cm}
	\par Regarding the complexity analysis of MCA, the time complexity is mainly determined by the time needed to explore the required RIS configurations. As outlined in Algorithm \ref{ag2}, MCA demonstrates a time complexity of $\mathcal{O}(T+2^{N_{\mathrm{new}}})$, directly linked to both the number of initially generated random configurations (i.e., $T$) and the number of unit configurations involved in cross-swapping (i.e., $N_{\mathrm{new}}$). The introduction of this method is primarily aimed at facilitating subsequent experimental measurement.
	\begin{figure}[tbp]
	\vspace{-0.4cm}
	\centering
	\includegraphics[height=1.6in]{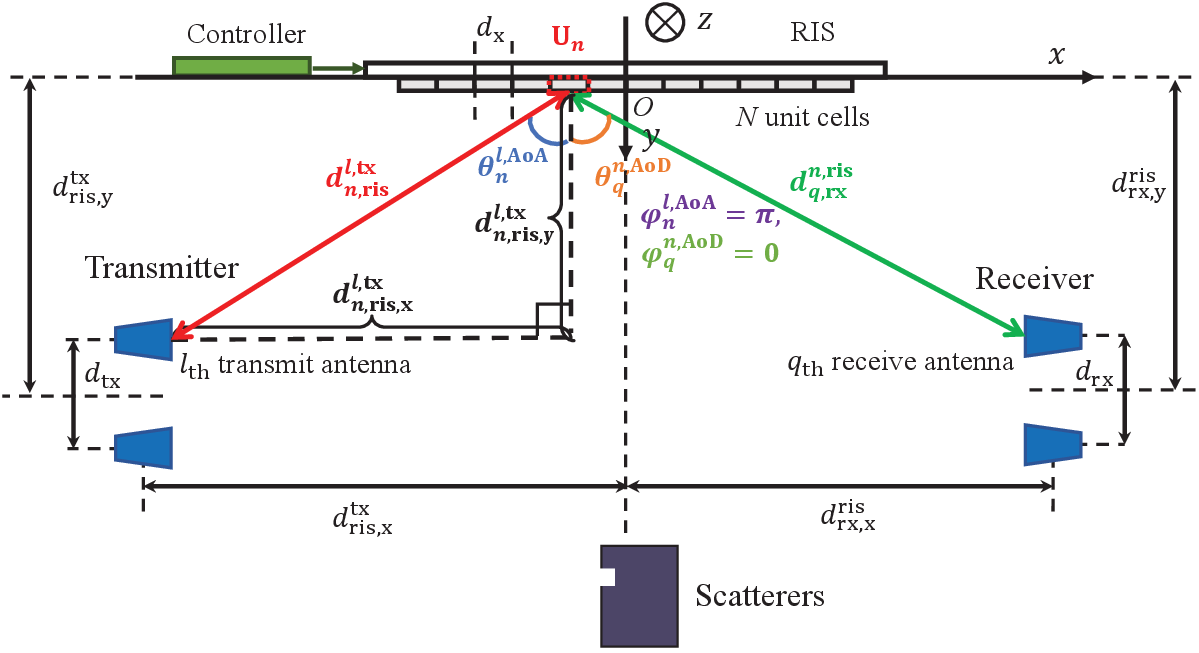}
	\vspace{-0.3cm}
	\caption{\raggedright Deployment of the considered RIS-assisted 2$\times$2 MIMO communication system (top view).}
	\label{Simmodel}
	\vspace{-0.5cm}
	\end{figure}

  \vspace{-0.3cm}
	\section{Numerical Simulation}
	In this section, we evaluate the performance improvement of the RIS-assisted MIMO channel through numerical simulations, utilizing the proposed optimization method. As illustrated in Fig. \ref{Simmodel}, we consider the column-controlled RIS, which can be treated as a uniform linear array (ULA). We set the centers of the transmit antennas, the receive antennas, and the RIS all positioned at the same height, thus reducing the system to two-dimension. The AoA $\varphi_{n}^{l, \mathrm{AoA}}$ and AoD $\varphi_{q}^{n, \mathrm{AoD}}$ are fixed at $\pi$ and $0$ in the x-y plane, respectively. In addition, $d_{\mathrm{ris,x}}^{\mathrm{tx}}$ ($d_{\mathrm{rx,x}}^{\mathrm{ris}}$) and  $d_{\mathrm{ris,y}}^{\mathrm{tx}}$ ($d_{\mathrm{rx,y}}^{\mathrm{ris}}$) represent the corresponding distance between the center of the transmitter (receiver) and the center of the RIS in the x or y direction, respectively. These distances are illustrated in Table \ref{tab1}. Moreover, $d_{\mathrm{tx}}$ and $d_{\mathrm{rx}}$ denote the transmit antenna spacing and the receive antenna spacing, respectively. When the transmit antennas and receive antennas are fixed, we can calculate $d_{n,\mathrm{ris}}^{l,\mathrm{tx}}$, $d_{q,\mathrm{rx}}^{n,\mathrm{ris}}$, $d_{q,\mathrm{rx}}^{l, \mathrm{tx}}$,  $d^{\mathrm{tx}}_{\mathrm{rx}}$, $\theta_{n}^{l, \mathrm{AoA}}$, and $\theta_{q}^{n, \mathrm{AoD}}$ based on the geometric relationship.
	
	\par The system operates at $f = 2.7$ GHz. Each column of the RIS can be seen as a macro unit cell, whose width and length are $d_{\mathrm{x}} = 0.05$ m and $d_{\mathrm{z}} = 1.6$ m, respectively. As shown in Fig. \ref{Simmodel}, the transmitter and receiver are both equipped with 2 antennas, that is $L = Q = 2$, with a transmit and receive gain being $G_{\mathrm{tx}} = G_{\mathrm{rx}} = 7$ dBi. More specific system parameters are listed in Table \ref{tab1}.
	
	\par When zero-forcing (ZF) equalization is applied to recover the data streams, the achievable system rate can be expressed as
	 \cite{matthaiou_mimo_2011}
	 \vspace{-0.3cm}
	\begin{equation}\label{equ10}
		R_{\mathrm{ZF}}\left( \mathbf{H},\rho \right) =\sum_{i=1}^{\mathrm{rank}(\mathbf{H})}{\log _2\left( 1+\frac{\rho}{\left[ \left( \mathbf{H}^H\mathbf{H} \right) ^{-1} \right] _{ii}} \right)},
		\vspace{-0.1cm}
	\end{equation}
	where $\rho$ is the transmit signal-to-noise ratio (SNR) and $\left[ \cdot \right] _{ii} $ denotes the $i_{\mathrm{th}}$ diagonal element of a matrix. In the high SNR regime where water filling power allocation reduces to equal power allocation, the capacity of the MIMO channel, which is introduced as the upper bound of the simulation, can be approximated by \cite{goldsmith_wireless_2005}
	\vspace{-0.3cm}
		\begin{equation}\label{equ11}
		C\left( \mathbf{H},\rho \right) =\sum_{i=1}^{\mathrm{rank}(\mathbf{H})}{\log _2\left( 1+{\sigma _i}^2\rho \right)}.
		\vspace{-0.2cm}
	\end{equation}
	\par In the following simulated cases, we set the transmit SNR $\rho = 50$ dB and Rician factor $K = 20$ dB. All the results are averaged over 1000 Monte Carlo channel realizations. We perform the MCA with $T = $160 and $N_{\mathrm{new}} =$ 4.
	
	\begin{table}[tbp]
		\vspace{-0.1cm}
		\centering
		\caption{PARAMETERS FOR THE NUMERICAL SIMULATION}
		\begin{center}
			\begin{tabular}{|c|c|c|}
				\hline
				\multirow{5}{*}{RIS Parameters}      
				& Operating Frequency $f$                                                                                                        & 2.7 GHz              \\ \cline{2-3} 
				& Coding Type                                                                                                                    & 1-bit phase         \\ \cline{2-3} 
				& Macro Unit Cell Number $N$                                                                                                     & 16/32/64            \\ \cline{2-3} 
				& Macro Unit Cell Width $d_{\mathrm{x}}$                                                                    & 0.05 m               \\ \cline{2-3} 
				& Macro Unit Cell Length $d_{\mathrm{z}}$                                                                    & 1.6 m          		\\ \hline
				\multirow{4}{*}{Distance Parameters} 
				&  Transmitter to RIS 
				$d_{\mathrm{ris,y}}^{\mathrm{tx}}$, $d_{\mathrm{ris,x}}^{\mathrm{tx}}$ 
				& 1 m, 0.5 m          \\ \cline{2-3} 
				& RIS to Receiver
				$d_{\mathrm{rx,y}}^{\mathrm{ris}}$, $d_{\mathrm{rx,x}}^{\mathrm{ris}}$    
				& 1 m, 0.5 m          \\ \cline{2-3} 
				& Transmit Antenna Spacing 
				$d_{\mathrm{tx}}$                                                                  & 0.1 m               \\ \cline{2-3} 
				& Receive Antenna Spacing $d_{\mathrm{rx}}$                                                                                      & 0.1 m               \\ \hline
			\end{tabular}
			\label{tab1}
		\end{center}
		\vspace{-0.3cm}
	\end{table}
	
	\begin{figure}[tbp]
		\vspace{-0.2cm}
		\centering
		\includegraphics[height=1.95in]{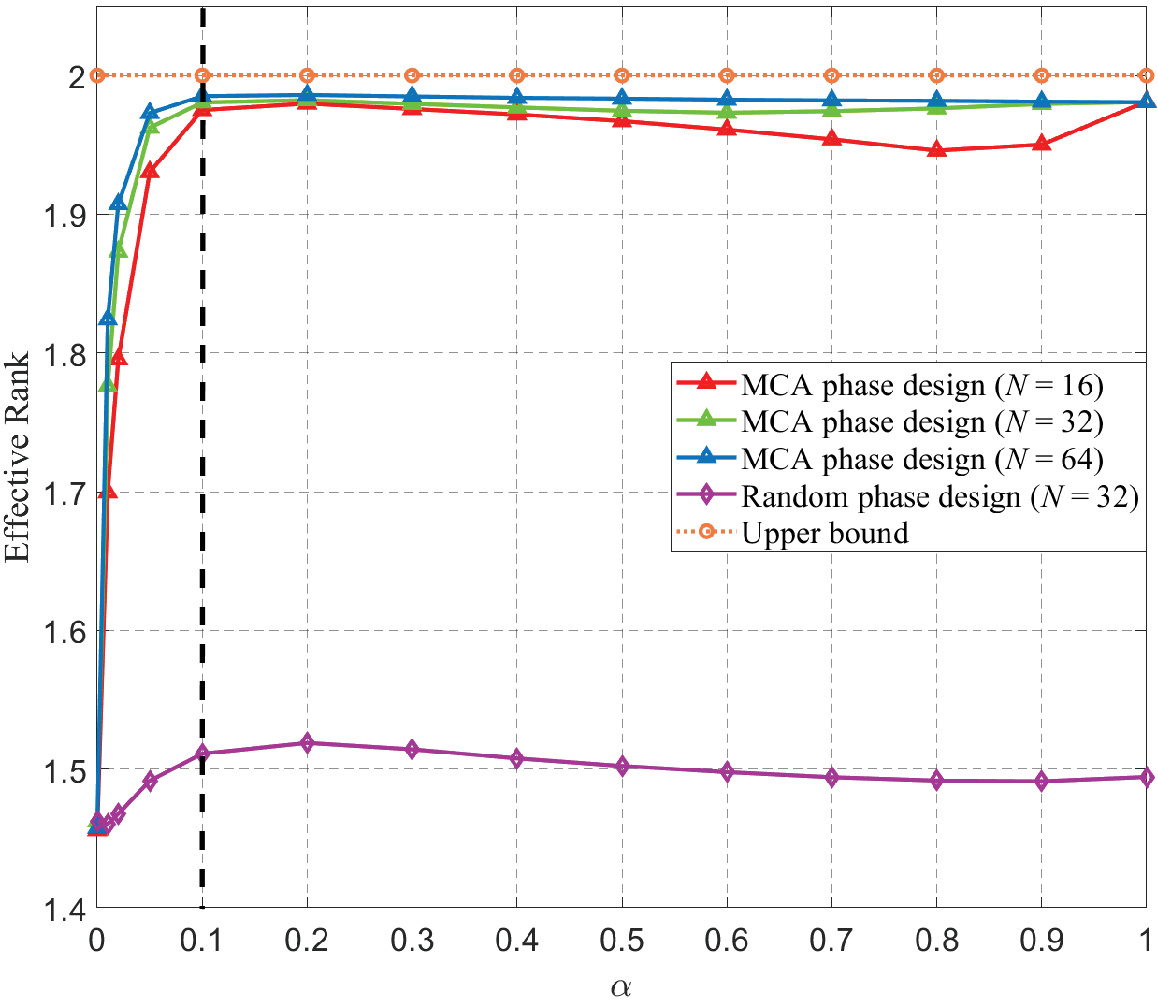}
		\vspace{-0.3cm}
		\caption{\raggedright Effective rank versus power ratio of the RIS-assisted channel for different numbers of unit cells.}
		\label{Erankresult}
		\vspace{-0.6cm}
	\end{figure}  
	  
	\begin{figure}[tp]
		\vspace{-0.2cm}
		\centering
		\includegraphics[height=1.95in]{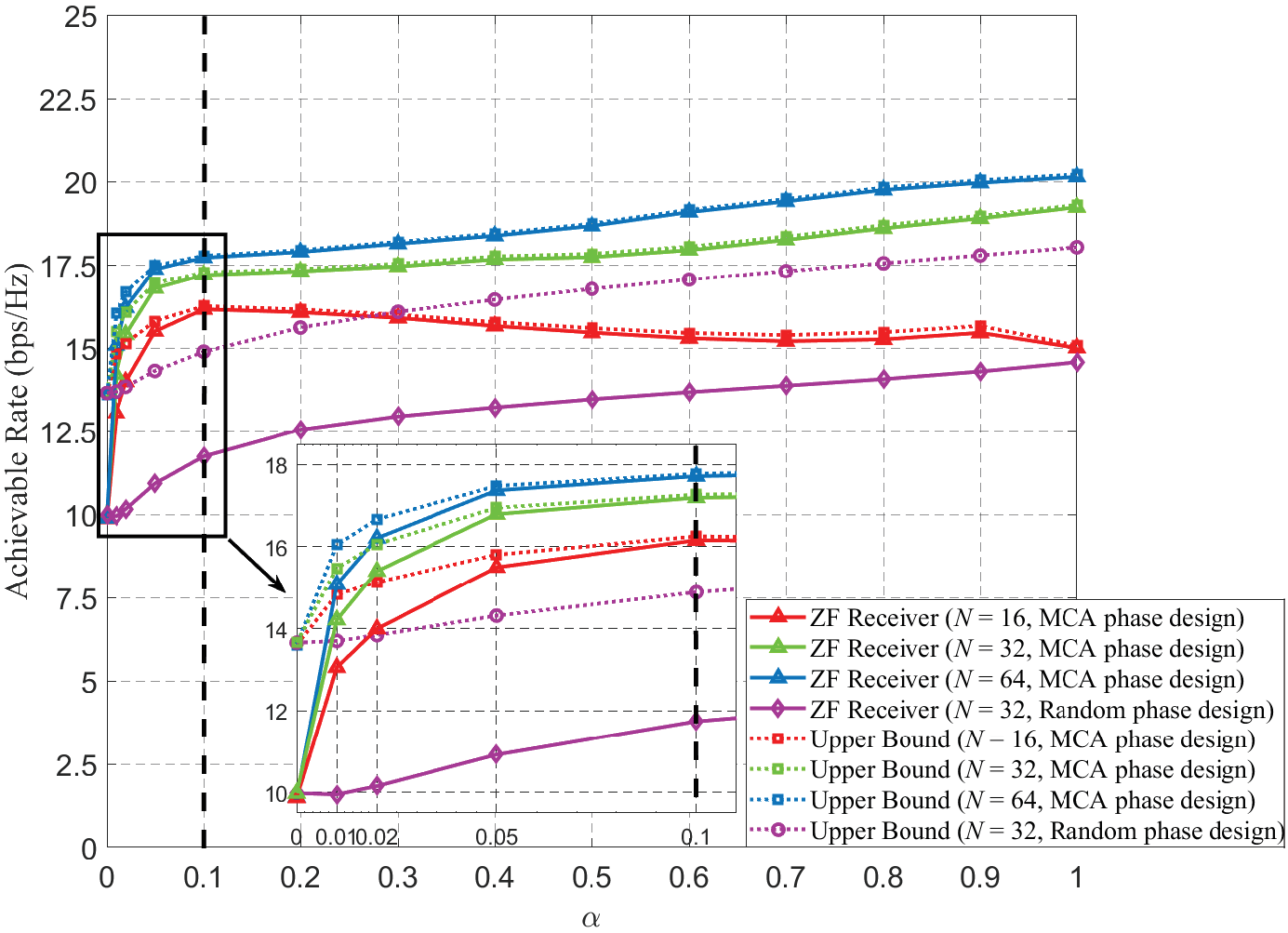}
		\vspace{-0.3cm}
		\caption{\raggedright Comparison of the achievable rate for the ZF receiver and the upper bound with increasing $\alpha$ when considering different numbers of unit cells.}
		\label{Acharesult}
		\vspace{-0.4 cm}
	\end{figure}
	
	\par Fig. \ref{Erankresult} demonstrates the effective rank versus power ratio of the RIS-assisted channel for different numbers of unit cells. When the power ratio $\alpha$ = 0, the MIMO channel $\mathbf{H}$ is only composed of the non-RIS-assisted channel, and the corresponding effective rank is 1.46. The reason for the deficient effective rank is that the MIMO channel exhibits a LoS-dominated characteristic when the Rician factor $K$ is set to 20 dB. Additionally, since the transmitter and receiver are relatively close to each other, the wavefront of the signals from the transmitter cannot be considered entirely planar when they reach the receiver. Therefore, the effective rank is slightly higher than the lower bound of 1. As $\alpha$ increases, the corresponding power ratio of the RIS-assisted channel $\mathbf{H}_{\mathrm{rx}}^{\mathrm{ris}}\mathbf{\Gamma}\mathbf{H}_{\mathrm{ris}}^{\mathrm{tx}}$ in the MIMO channel $\mathbf{H}$ rises. At this point, when utilizing the RIS optimized by MCA, the effective rank of matrix $\mathbf{H}$ quickly approaches the upper bound of 2. However, the enhancement in the effective rank observed in the MIMO channel with the RIS configured with random phase is minimal, approximately 0.06. The results above reveal that leveraging the well-configured RIS can improve the effective rank of the MIMO channel.
	
	\par We also evaluate the achievable rate of the ZF receiver and compare it with the upper bound as $\alpha$ increases, while considering varying numbers of unit cells. As shown in Fig. \ref{Acharesult}, when $N = $ 16, the achievable rate of the ZF receiver increases first and then decreases. The upward trend is mainly attributed to the improvement of the effective rank of MIMO channel $\mathbf{H}$. The subsequent downward trend occurs due to the limited number of unit cells, as the gain provided by the RIS-assisted channel is insufficient to compensate for the losses in the non-RIS-assisted channel. Thus, when power ratio $\alpha$ of the RIS-assisted channel $\mathbf{H}_{\mathrm{rx}}^{\mathrm{ris}}\mathbf{\Gamma}\mathbf{H}_{\mathrm{ris}}^{\mathrm{tx}}$ in the channel $\mathbf{H}$ increases, the corresponding achievable rate actually decreases. However, when the RIS consists of more unit cells, i.e., $N =$ 32 or 64, the RIS-assisted channel has higher gains. As a result, the achievable rate of the ZF receiver continues to increase. Increasing $N$ enhances the achievable rate of the ZF receiver. This effect becomes more pronounced with larger $\alpha$.
	
	\par Additionally, it can be observed that as the power ratio $\alpha$ increases, the effective rank of the 2$\times$2 MIMO channel $\mathbf{H}$ approaches the upper limit of 2 in the right region indicated by the black dashed line in Fig. \ref{Erankresult}. This result implies that the two subchannels of $\mathbf{H}$ are orthogonal to each other. Meanwhile, the achievable rate of the simple ZF receiver approaches the upper bound, as shown in Fig. \ref{Acharesult}. However, for the system assisted by the RIS configured with random phase, the effective rank of the corresponding MIMO channel doesn't approach the upper limit. Consequently, the achievable rate of the simple ZF receiver falls far short of the upper bound. Leveraging the well-configured RIS to optimize the effective rank of the MIMO channel, we can enhance the MIMO channel itself, simplify the receiver algorithm, and ultimately improve the overall performance of the system.

	\section{Experimental Measurement}
\begin{figure}[bp]
		\vspace{-0.6cm}
		\centering
		\includegraphics[height=1.9in]{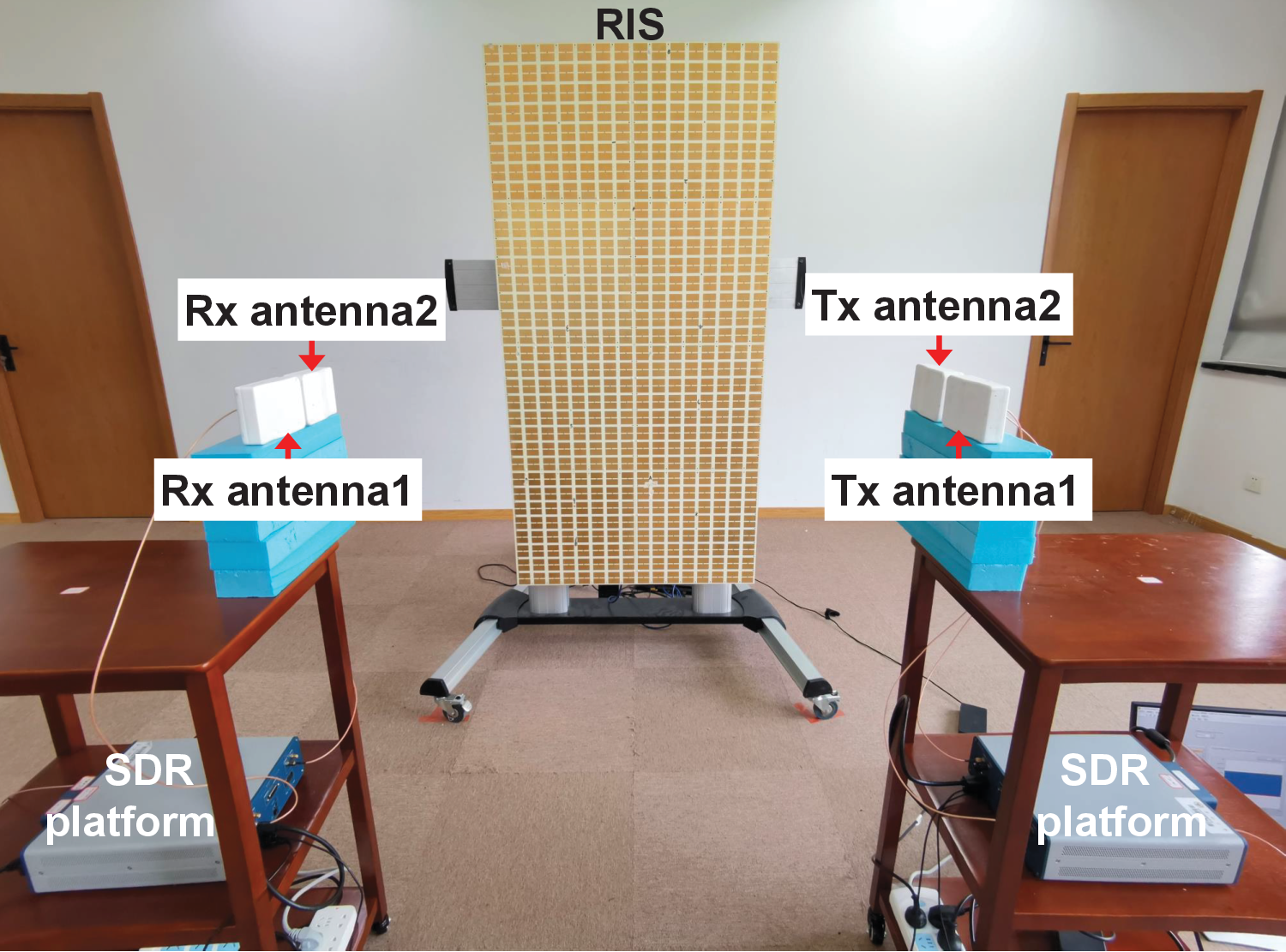}
		\vspace{-0.3cm}
		\caption{\centering A photo of the RIS-assisted 2$\times$2 MIMO communication prototype.}
		\label{Expesystem}
		\vspace{-0.6cm}
	\end{figure}

	In this section, we set up an RIS-assisted 2$\times$2 MIMO communication prototype to conduct experimental measurements. As depicted in Fig. \ref{Expesystem}, the transmitter and the receiver consist of two antennas and the software defined radio (SDR) platform, respectively. The RIS is placed aside, with its center aligned with the center of all antennas at the same height, thereby establishing an RIS-assisted 2×2 MIMO communication system. The parameters of the fabricated RIS and the corresponding distances are identical to those utilized in the numerical simulation.
	
	\par Upon the configuration of RIS is completed, the receiver employs the least squares (LS) method to estimate the overall cascaded MIMO channel $\mathbf{H}$ based on pilot signals transmitted from the transmitter. Subsequently, the corresponding effective rank of the channel is calculated based on equation (\ref{equ7}) as a metric for the MCA to optimize the RIS configuration. The achievable rate of the ZF receiver can be obtained by summing up the achievable rates of each independent data stream after ZF equalization, which can be expressed as
	\vspace{-0.1 cm}
	\begin{equation}
		R_{\mathrm{ZF}}^{\mathrm{mea}}\left( \mathbf{H},\rho _{i}^{\mathrm{ZF}} \right) =\sum_{i=1}^{\mathrm{rank}\left( \mathbf{H} \right)}{\log _2\left( 1+\rho _{i}^{\mathrm{ZF}} \right)},
	\vspace{-0.1cm} 
	\end{equation}
	where $\rho_{i}^{\mathrm{ZF}}$ is the SNR of the $i_{\mathrm{th}}$ data stream after ZF equalization. The specific measurement method for each SNR $\rho_{i}^{\mathrm{ZF}}$ can be found in Sections IV-D and V-B of reference \cite{tang_mimo_2020}. 
	
	\par
\begin{figure}[tbp]
		\centering
		\includegraphics[height=2in]{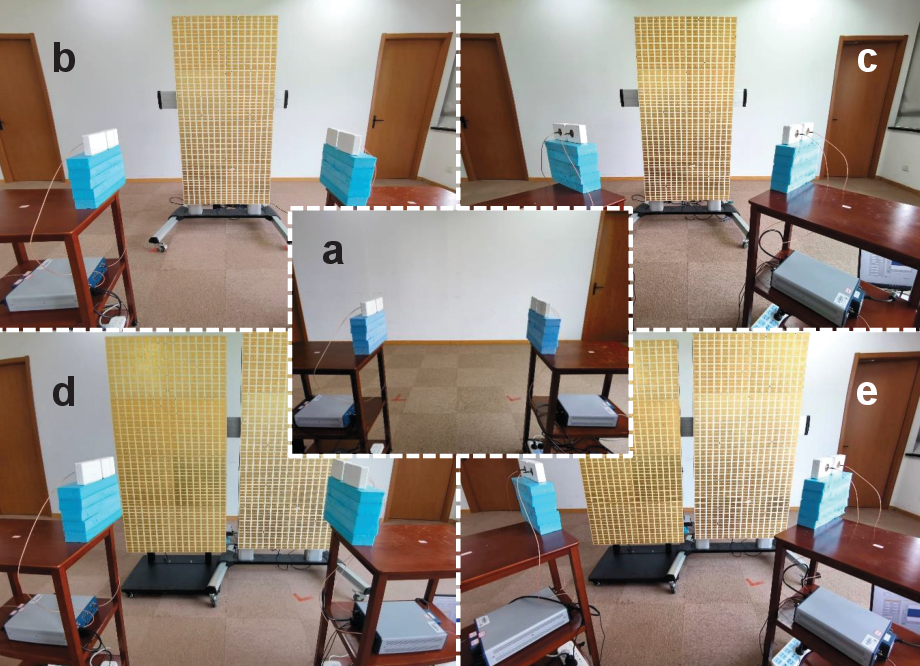}
			\vspace{-0.3cm}
		\caption{\raggedright Layout of experimental scenarios: (a) Transmitter faces receiver without RIS, (b) Transmitter faces receiver with one RIS aside, (c) Transmitter and receiver both face one RIS, (d) Transmitter faces receiver with two RISs aside, (e) Transmitter and receiver both face two RISs.}
		\label{Expedeployment}
		\vspace{-0.6cm}
	\end{figure}

 Since the transmit and receive antennas we employed are directional antennas with a gain of 7 dBi, we can modify the power ratio $\alpha$ of the RIS-assisted channel $\mathbf{H}_{\mathrm{rx}}^{\mathrm{ris}}\mathbf{\Gamma}\mathbf{H}_{\mathrm{ris}}^{\mathrm{tx}}$ in the MIMO channel $\mathbf{H}$ by altering the orientation of the antennas towards the RISs. As depicted in Fig. \ref{Expedeployment}(a), when the transmitter directly faces the receiver without any RIS, the corresponding $\alpha$ value is 0. In Fig. \ref{Expedeployment}(b) and Fig. \ref{Expedeployment}(d), where the transmitter faces the receiver with the RISs placed aside, the corresponding $\alpha$ slightly exceeds 0. As shown in Fig. \ref{Expedeployment}(c) and Fig. \ref{Expedeployment}(e), when both the transmitter and the receiver face the RISs, the corresponding $\alpha$ approaches 1. Furthermore, we evaluated the impact of increasing the number of unit cells by comparing the results obtained with one RIS and two RISs. Table \ref{tab2} summarizes the corresponding measurement results.
	
	\par It can be observed that when using one RIS ($N$ = 16), the achievable rate of the ZF receiver $R_{\mathrm{ZF}}^{\mathrm{mea}}$ increases from 17.7 to 19.4 bps/Hz at the beginning and then decreases to 18.9 bps/Hz as $\alpha$ increases. Meanwhile, the effective rank rises from 1.53 to 1.96.  However, when using two RISs ($N$ = 32), $R_{\mathrm{ZF}}^{\mathrm{mea}}$ continuously increases by 13.6\%, from 17.7 to 20.1 bps/Hz as $\alpha$ increases. Simultaneously, the effective rank rises from 1.53 to 1.96, obtaining a 30.1\% increase. In scenarios where the RIS-assisted channel dominates $\mathbf{H}$, as depicted in Fig. \ref{Expedeployment}(c) and Fig. \ref{Expedeployment}(e), adding one RIS (increasing $N$ from 16 to 32) leads to a 1.2 bps/Hz increase in $R_{\mathrm{ZF}}^{\mathrm{mea}}$. The measurement results are consistent with the simulation results and verify that RIS can be effectively used to optimize the effective rank of the channel and improve the system performance.
	
	\begin{table}[tbp]
	\vspace{-0.5 cm}
		\centering
		\caption{RESULTS OF THE EXPERIMENTAL MEASUREMENT}
		\vspace{-0.5cm}
		\begin{center}
			\begin{tabular}{|c|c|c|c|c|c|}
				\hline
				No. &  $\alpha$ and $ N$                                                       & $\mathrm{erank}(\mathbf{H})$&\begin{tabular}[c]{@{}c@{}}$\rho_{1}^{\mathrm{ZF}}$\\(dB)\end{tabular}   &\begin{tabular}[c]{@{}c@{}}$\rho_{2}^{\mathrm{ZF}}$\\(dB)\end{tabular}  &\begin{tabular}[c]{@{}c@{}} $R_{\mathrm{ZF}}^{\mathrm{mea}}$\\(bps/Hz)\end{tabular} \\ \hline
				a             & $\alpha = 0$                                                                       & 1.53           & 26.6                       & 27.1                       & 17.7                                 \\ \hline
				b            &\begin{tabular}[c]{@{}c@{}}$\alpha$ slightly exceed 0\\ $N$ =     16\end{tabular} & 1.70           & 27.9                       & 30.2                       & 19.4                                 \\ \hline
				c             &\begin{tabular}[c]{@{}c@{}}$\alpha$ approaches 1\\ $N$ =     16\end{tabular}           & 1.96           & 28.1                       & 28.8                       & 18.9                                 \\ \hline
				d             &\begin{tabular}[c]{@{}c@{}}$\alpha$ slightly exceed 0\\ $N$ =     32\end{tabular} & 1.65           & 28.1                       & 29.5                       & 19.2                                 \\ \hline
				e              &\begin{tabular}[c]{@{}c@{}}$\alpha$ approaches 1\\ $N$ =     32\end{tabular}           & 1.99           & 30.2                       & 30.4                       & 20.1                                 \\ \hline
			\end{tabular}
		\end{center}
		\label{tab2}
		\vspace{-0.8cm}
	\end{table}
	\vspace{-0.2cm}
	\section{Conclusion}
	In this study, we proposed an efficient RIS optimization method to enhance the effective rank of the MIMO channel for achievable rate improvement in RIS-assisted systems. Through numerical simulations and experimental measurements, we have confirmed that RISs can be utilized to improve the effective rank of the MIMO channel as well as the achievable rate effectively. The results show that the RIS-assisted MIMO channel optimized using MCA, compared to the scenario without RIS, exhibited an increase of 30.1\% in effective rank and a 13.6\% achievable rate improvement. Based on these promising findings, our study reveals the superior performance of RIS in optimizing MIMO channels, which highlights the potential applications of RISs in future networks. Furthermore, in our upcoming research efforts, we will continue to investigate this topic in depth, such as studying the impact of the number of bits in the RIS configuration on the performance of the channel optimization, and so on.

\end{document}